\documentclass[prl,aps,epsfig,twocolumn]{revtex4}
\usepackage{amssymb}
\usepackage{amsmath}
\usepackage{amsfonts}
\usepackage{graphicx}

\providecommand{\U}[1]{\protect\rule{.1in}{.1in}}

\begin{document}

\textbf{Comment on \textquotedblleft Unified Formalism of Andreev Reflection
at a Ferromagnet/Superconductor Interface\textquotedblright\ by T.~Y. Chen,
Z. Tesanovic, and C.~L. Chien}

A recent paper of Chen \textit{et} \textit{al}.~\cite{Chen}, claims to have
derived an allegedly, previously unavailable \textquotedblleft
unified\textquotedblright\ Andreev Reflection (AR) formalism for an
arbitrary spin polarization $P$ that recovers earlier results as its special
limits. In this Comment we show that, contrary to this claim, there are
numerous works correctly solving the problem formulated in Ref.~\cite{Chen}
for an arbitrary $P$ \cite{deJong,Mazin01,Kashiwaya,Zutic99,Chalsani,Eschrig}%
, while Ref.~\cite{Chen} fails to correctly incorporate $P\neq0$ effects and
violate basic physical principles.

In the $P=0$ limit the approach of Ref.~\cite{Chen} is identical to the one
dimensional (1D) BTK model~\cite{Blonder} for nonmagnetic metals, while for $%
P\neq0$ the only difference is postulating an AR wavefunction that has an
additional evanescent wave contribution parameterized by a decay constant $%
\alpha$, $\psi_{AR}=a%
\begin{Bmatrix}
0 \\ 
1%
\end{Bmatrix}
\exp[(\alpha+i)q^{+}x]$, with $\alpha=2\sqrt{P/(1-P)}$. However, this
wavefunction violates the charge conservation: the corresponding charge
current density is $j_{AR}(x)\propto\mathrm{Im}[\psi_{AR}^{\ast}\nabla%
\psi_{AR}]\propto q^{+}|a|^{2}\exp[2\alpha q^{+}x]$. In the F-region, for
any $0<\alpha<\infty$, \textit{i.e.}, $0<P<1$, the divergence of the \emph{%
total} current is finite and thus the total charge is not conserved (in
particular, for $0<x\lesssim{1}/{2\alpha q^+}$), signaling that $\psi_{AR}$
is unphysical. Even for the $P=1$ half-metal (HM) state, where the above
expression correctly gives $j_{AR}=0,$ it is still incorrect. For a HM the
AR electron decays with a finite (not infinitely small, as postulated in Ref.%
\cite{Chen})  penetration depth, which depends on the electronic structure
details, primarily the size of the gap in the nonmetallic spin channel~\cite%
{note1}.

The rationale for inventing a new $\psi_{AR}$~\cite{Chen} was based on the
premise that: (a) the result in Ref.~\cite{Mazin01} was derived only in the
extreme case of a fully-polarized HM with $\tilde{\psi }_{AR}=a%
\begin{Bmatrix}
0 \\ 
1%
\end{Bmatrix}
\exp(\kappa x$) and a nonmagnetic metal with $\tilde{\psi}_{AR}=a%
\begin{Bmatrix}
0 \\ 
1%
\end{Bmatrix}
\exp(ikx$), while postulating that for an intermediate $P$ the current will
be a linear combination of these cases, (b) no prior work had treated a $%
0<P<1$ case, and (c) one can meaningfully define $P$ of an individual
electron. Regarding (a-b),  the derivation in Ref.~\cite{Mazin01}, as well
as in other works~\cite{Kashiwaya,Zutic99,Chalsani,Eschrig}, is rigorous for
an arbitrary $P$. The last point, (c) is particularly misleading. In fact,
one cannot define a BTK model in 1D with an arbitrary $P$ because any given
electron in a metal Andreev-reflects either into a propagating, or into an
evanescent wave. Finite $P$ simply means that some current-carrying
electrons belong to the former group and the others to the latter; one can
only define transport spin polarization in a multielectron system, where the
numbers of electron states at the Fermi surface (conductivity channels) for
the two spin directions differ. If the electron wavevector $\mathbf{k}$ is
decomposed into a non-conserved $k_{x}$ (normal to the F/S interface) and
conserved~\cite{note2} 2D $\mathbf{{k}_{\Vert}}$, then, after the usual
quantization of $\mathbf{{k}_{\parallel},}$ one finds that the total number
of states available for transport in the $x$ direction (\textit{i.e., }the
number of the conductivity channels) for a given spin direction is
proportional to the total area of the Fermi surface projection on the
interface plane, given by the Fermi surface average of the Fermi velocity, $%
n=\left\langle N(E_{F})v_{Fx}\right\rangle $~\cite{Mazin01}. After summation
over $all$ electronic states the total current $turns$ $out$ to be a linear
combination (with the weights defined by $P),$ of the solutions of the 1D
BTK model with $P=0$ and $P=1.$ This was not postulated, but derived, in
numerous papers (see the discussion in Ref.~\onlinecite{Eschrig}).

In contrast to the \textquotedblleft universal\textquotedblright\ $P$ in the
Ref.~\cite{Chen}, independent of electronic mass, Fermi wavevectors, or any
band structure at all, the real transport spin polarization for AR
spectroscopy depends on the \emph{overall} Fermi surface properties.
Moreover, there is no unique spin polarization (even for a uniform bulk
metal), as it depends on an experimental probe. In fact, the definition used
in Ref.~\cite{Chen} (neglecting the Fermi velocities) does not correspond to
the AR experiments, but rather to spin-polarized photoemission.

To summarize, Ref.~\cite{Chen} has misinterpreted or ignored previous works
where the posed problem has been correctly solved for an arbitrary $P,$ and
attempted to supplant the existing solution with an incorrect formula,
postulating an unphysical wavefunction for the reflected electron, which
does not conserve charge and has an incorrect HM limit. They proceeded by
calculating the current due to $\psi _{AR}$ at $x=0,$ even though the actual
current is measured far away from the interface (where they would have
gotten zero Andreev contribution for any $P$ and grave disagreement with the 
experiment). Furthermore, they have arbitrarily decided that the
penetration depth for an electron inside the band gap in the
transport-spin-minority channel is uniquely defined by the spin polarization
(these two quantities are in fact unrelated). As a result, they arrived at a
formula that for their own experimental data provides a fit that is
essentially identical to that obtained by using Ref.~\cite{Mazin01} (the
difference is below potential errors introduced by the BTK approximations of
a step-shaped pairing potential and spin-independent $\delta $-shaped
barrier~\cite{note3}).

Finally, we note that the inclusion of inelastic scattering using a
phenomenological finite $\Gamma\neq0$, another claimed novelty of Ref.~\cite%
{Chen} has already been previously employed for AR with arbitrary $P$~\cite%
{Chalsani,Cohen}.

It may be that the formulas contrived in Ref. \cite{Chen} fit a particular experiment.
 However, there is a maxim
attributed to Niels Bohr, that there exists an infinite number of incorrect theories that
 correctly describe the finite number of experiments. 

M. Eschrig$^1$, A.~A. Golubov$^{2}$, I.~I. Mazin$^{3}$,  B. Nadgorny$^{4}$, 
Y. Tanaka$^5$, O.~T. Valls$^{6}$, Igor \v{Z}uti\'c$^{7}$ \newline
{\footnotesize $^1$ Department of Physics, Royal Holloway, University of
London, Egham, Surrey TW20 0EX, UK,  $^{2}$Faculty of Science and
Technology and MESA+ Institute of Nanotechnology, University of Twente, The
Netherlands, $^{3}$Naval Research Laboratory, Washington, DC 20375, USA,  $^4
$Department of Physics and Astronomy, Wayne State University, Detroit, MI
48201, USA, $^5$Department of Applied Physics, Nagoya University, Aichi
464-8603, Japan, $^{6}$School of Physics and Astronomy, University of
Minnesota, Minneapolis, MN 55455, USA, $^{7}$Department of Physics,
University at Buffalo, SUNY, Buffalo, NY 14260, USA }


\begin{thebibliography}{99}
\bibitem{Chen} T. Y. Chen, Z. Tesanovic, and C. L. Chien, Phys. Rev. Lett. 
\textbf{109}, 146602 (2012).

\bibitem{deJong} M.~J.~M. de Jong and C.~W.~J. Beenakker, Phys. Rev. Lett. 
\textbf{74}, 1657 (1995).

\bibitem{Mazin01} I. I. Mazin, A. A. Golubov, and B. Nadgorny, J. Appl.
Phys. \textbf{89}, 7576 (2001); G. T. Woods, R. J. Soulen, Jr., I. I. Mazin,
B. Nadgorny, M. S. Osofsky, J. Sanders, H. Srikanth, W. F. Egelhoff, and R.
Datla, Phys. Rev. B \textbf{70}, 054416 (2004).

\bibitem{Kashiwaya} S. Kashiwaya, Y. Tanaka, N. Yoshida, and M. R. Beasley,
Phys. Rev. B \textbf{60}, 3572 (1999).

\bibitem{Zutic99} I. \v{Z}uti\'c and O.~T. Valls, Phys. Rev. B \textbf{61},
1555 (2000), see Eq.~(2.8); I. \v{Z}uti\'c and S. Das Sarma, Phys. Rev. B 
\textbf{60}, R16322 (1999).

\bibitem{Chalsani} P. Chalsani, S.~K. Upadhyay, O. Ozatay, and R.~A.
Buhrman, Phys. Rev. B \textbf{75}, 094417 (2007).

\bibitem{Eschrig} R. Grein, T. Lofwander, G. Metalidis, and M. Eschrig,
Phys. Rev. B \textbf{81}, 094508 (2010).

\bibitem{Blonder} G.~E. Blonder and M. Tinkham and T.~M. Klapwijk, Phys.
Rev. B \textbf{25}, 4515 (1982).

\bibitem{note1} Ref.~\cite{Mazin01}, as well as some others, eventually uses
the limit of the infinitely small penetration depth, because the final
result depends on this parameter very little, but it is important to use in
the derivation a physically meaningful wavefunction that allows for a finite
penetration depth. This inconsistency in Ref. ~\cite{Chen} is a part of a
bigger problem of using the same $k_F$ for both spin channels, despite the
fact that as $P \rightarrow 1$ one of the two $k_F^\pm$ gradually vanishes~%
\cite{deJong,Kashiwaya,Zutic99,Eschrig}.

\bibitem{note2} For a ballistic flat interface. The generalization onto a
diffusive case, where only the total number of the conductivity channels for
each spin direction, but not the individual $\mathbf{k}_{\parallel}$ are
conserved, is straightforward~\cite{Mazin01}.

\bibitem{note3} For other parameters [see their Fig. 2(c)] their approach
leads to unphysical kinks at zero bias and unphysical notches near the gap
voltage.

\bibitem{Cohen} Other authors observed that instead of adding a $\Gamma$, at
an extra computational cost, one can simply artificially increase the
temperature in the original expressions, since the effect of the two types
of broadening on the conductance curves is essentially the same, see, $e.g.$%
,Y. Bugoslavsky, Y. Miyoshi, S. K. Clowes, W. R. Branford, M. Lake, I.
Brown, A. D. Caplin, and L. F. Cohen, Phys. Rev. B \textbf{71}, 104523 (2005)
\end{thebibliography}
\end{document}